\documentclass[a4paper,11pt]{article}

\usepackage{pos}

\usepackage{amsmath}

\usepackage{xcolor}
\usepackage{tikz}
\usetikzlibrary{positioning}
\usepackage{youngtab}
\usepackage{ytableau}

\usepackage{comment}

\newcommand{\bea}{\begin{eqnarray}}
\newcommand{\eea}{\end{eqnarray}}

\newcommand{\mC}{\mathbb{C}}

\newcommand{\Vmn}{V_N^{\otimes m}\otimes \overline{V}_N^{\otimes n}}
\newcommand{\BRT}{{\rm BRT}}

\newcommand{\cR}{\mathcal{R}}
\newcommand{\cZ}{\mathcal{Z}}

\newcommand{\hght}{\operatorname{height}}

\newcommand{\MixedYoung}[4][0.5]{%
  \begingroup
  \def\cell{#1}%
  \def\nrows{#2}%
  \begin{tikzpicture}[x=\cell cm,y=\cell cm,baseline={(0,-0.5*\nrows)}]
    \draw[thick] (0,0) -- (0,-\nrows);
    \foreach \i in {1,...,#2} {
      \draw (-0.1,-\i) -- (0.1,-\i);
    }
    \foreach \r [count=\i from 1] in {#4} {%
      \ifnum\r>0
        \foreach \j in {0,...,\numexpr\r-1\relax} {%
          \draw (\j,-\i) rectangle (\j+1,-\i+1);
        }%
      \fi
    }
    \foreach \l [count=\i from 1] in {#3} {%
      \ifnum\l>0
        \foreach \j in {0,...,\numexpr\l-1\relax} {%
          \draw (-\j-1,-\i) rectangle (-\j,-\i+1);
        }%
      \fi
    }
  \end{tikzpicture}
  \endgroup
}

\title{Oscillators from non-semisimple walled Brauer algebras}

\author*[a]{Sanjaye Ramgoolam}
\author[b]{Micha\l~Studzi\'nski}

\affiliation[a]{Queen Mary University of London,\\
327 Mile End Road, London, United Kingdom}

\affiliation[b]{International Centre for Theory of Quantum Technologies,\\
University of Gda\'nsk, Poland}

\emailAdd{s.ramgoolam@qmul.ac.uk}
\emailAdd{michal.studzinski@ug.edu.pl}

\abstract{
The walled Brauer algebras $B_N(m,n)$ govern Schur--Weyl duality for unitary groups
$U(N)$ acting on mixed tensor spaces $V_N^{\otimes m}\otimes \overline{V}_N^{\otimes n}$ and
play an important role in applications ranging from AdS/CFT to quantum information theory.
In the stable regime $N\ge m+n$ the algebra is semisimple and its representation theory is well
understood. For $N<m+n$, however, $B_N(m,n)$ becomes non-semisimple. The representation of the
algebra on tensor space has a non-trivial kernel and the corresponding quotient algebra is
semisimple, with representation dimensions differing from those in the stable regime.

We introduce \emph{restricted Bratteli diagrams}, obtained by modifying the standard Bratteli
diagrams for $B_N(m,n)$. This construction provides a systematic way to use
representation-theoretic data from the stable regime to compute the dimension modifications
arising in the non-semisimple regime. In the regime $N=m+n-l$, with $l$ small compared to
$m,n$, we show that the restricted diagrams exhibit a stability property and enable an
efficient counting of the paths responsible for these dimension corrections.

Remarkably, the resulting generating functions are governed by the partition function of an
infinite tower of simple harmonic oscillators. We briefly discuss implications for the
construction of orthogonal bases of matrix invariants in gauge theory and related applications
in quantum information theory.
}

\FullConference{Proceedings of the Corfu Summer Institute 2025
``School and Workshops on Elementary Particle Physics and Gravity'' (CORFU2025)\\
27 April -- 28 September, 2025\\
Corfu, Greece}

\begin{document}
\maketitle

\section{Introduction}

This short proceedings article summarises the longer  paper \cite{SHOWBA}.

Schur--Weyl duality~\cite{FultonHarris1991,Howe1995} provides a powerful bridge between the representation theory of the classical Lie groups and that of diagram algebras acting on tensor spaces. For the action of the unitary
group $U(N)$ on tensor powers of the fundamental representation, the dual commutant algebra  is the group
algebra of the symmetric group. In the case of mixed tensor spaces
$V_N^{\otimes m}\otimes  \overline{V}_N^{\otimes n}$, the corresponding commutant is the walled
Brauer algebra $B_N(m,n)$~\cite{BCHLLS1994,Turaev1990, Koike1989,Benkart1996,Bulgakova2020,CDVM2008}. 
The dimensions of the irreducible representations of the commutant algebras are equal to the multiplicities of 
corresponding representations of the unitary group. 
The walled Brauer algebras exist at an interface of 
representation theory, gauge--string duality, and quantum information theory. In the
AdS/CFT context they underlie the construction of orthogonal bases of matrix invariants and
related multi-matrix observables \cite{KimuraRamgoolam2007,BhattacharyyaCollinsKoch2008,BhattacharyyaKochStephanou2008,BrownHeslopRamgoolam2008,BrownHeslopRamgoolam2009,KimuraRamgoolam2008,Kimura:2009jf,Kimura2010,KimuraLin2012,MattioliRamgoolam2016}, while in quantum information they arise in mixed
Schur--Weyl duality~\cite{StudzinskiMlynikMozrzymasHorodecki2025,GrinkoBurchardtOzols2023,MozrzymasStudzinskiHorodecki2018}, teleportation protocols \cite{
GrinkoBurchardtOzols2024,FeiTimmermanHayden2023,WillsHsiehStrelchuk2024,
StudzinskiStrelchukMozrzymasHorodecki2017,StudzinskiMozrzymasKopszak2021}, and other settings where symmetry reduction
plays a central role~\cite{Grinko2025,Nguyen2023,QuintinoEbler2022,GrinkoOzols2024,CerveroMartinMancinskaTheil2024,EggelingWerner2001}.

A particularly attractive feature of the walled Brauer algebra is the existence of a stable,
or large-\(N\), regime. When $N\ge m+n$, the algebra $B_N(m,n)$ is semisimple and its
irreducible representations are labelled by Brauer representation  triples (BRT) which are reviewed  in Section \ref{sec:SWduality}. The dimensions of these representations  are given by
explicit combinatorial formulae independent of $N$. Equivalently, the decomposition of mixed
tensor space into irreducible representations of $U(N)$ exhibits a stability property: the
multiplicities are controlled by representation-theoretic data of the semisimple Brauer
algebra and can be computed in terms of path counting in Bratteli diagrams. 

For $N<m+n$, however, the situation changes substantially. The algebra $B_N(m,n)$ becomes
non-semisimple, the natural representation on tensor space develops a non-trivial kernel,
and the relevant object on the commutant side is a semisimple quotient algebra. In this
regime, some admissible Brauer triples still retain the same dimensions as in the stable
regime, while others acquire non-trivial corrections. Understanding these modified
dimensions is therefore a natural and non-trivial combinatorial problem. It is also relevant
for applications, since these multiplicities enter both the structure of gauge-invariant
operator bases and the symmetry-adapted analysis of quantum-information protocols. 

In the paper \cite{SHOWBA} we defined \emph{restricted Bratteli diagrams} (RBD), obtained by refining
the usual Bratteli diagrams so as to isolate the data responsible for dimension
corrections in the non-semisimple regime. This provides a systematic way to use
representation-theoretic information from the stable regime in order to compute modified
dimensions for the quotient algebra. Focusing on the regime $N=m+n-l$ with fixed positive  $l$, we
show that these restricted diagrams display a new stability property for sufficiently large
$m$ and $n$, and that their combinatorics is controlled by generating functions closely
related to the partition function of an infinite tower of simple harmonic oscillators.

The paper is organised as follows. We begin by reviewing the representation-theoretic
background for mixed tensor Schur--Weyl duality and the role of the walled Brauer algebra
\(B_N(m,n)\) as the commutant of the \(U(N)\) action on
\(V_N^{\otimes m}\otimes \overline{V}_N^{\otimes n}\). In particular, we recall the stable
semisimple regime \(N\ge m+n\), where irreducible representations are labelled by Brauer
triples and their dimensions are computed by standard large-\(N\) formulae, together with
the Bratteli-diagram interpretation of these dimensions in terms of path counting. 

We then turn to the non-semisimple regime \(N<m+n\), where the natural action of
\(B_N(m,n)\) on tensor space has a non-trivial kernel and taking a quotient by the kernel gives a
semisimple algebra \(\widehat{B}_N(m,n)\) which is the commutant of the unitary group in mixed tensor space. 
We describe the restricted
Bratteli diagrams (RBD), which are simplified versions of the full Bratteli-diagram
construction adapted to the non-semisimple regime. The RBD have nodes of two colors, which we describe as red and green.

Next, we study the regime \(N=m+n-l\) with \(l\) fixed and small compared to \(m\) and
\(n\). In this setting, the RBD exhibit a new stability phenomenon: once
\(m,n\ge 2l-3\), their structure becomes independent of the particular values of \(m\) and
\(n\), depending only on \(l\). This \((m,n)\)-stability allows one to separate the
universal combinatorics of the non-semisimple corrections from the details of the ambient
mixed tensor space, and it leads to effective formulae for modified dimensions in the
cases \(l=2,3,4\). 

Having established the structural properties of the restricted diagrams, we then study
the problem of  counting the red and green nodes in these diagrams as a function of the depth. We find, surprisingly, 
 that these counting functions are governed by a universal generating function: 
\bea\label{eq:ZunivIntro}
\cZ_{\rm univ}(x)=\frac{x}{(1-x)(1-x^2)}\prod_{i=1}^{\infty}\frac{1}{(1-x^i)^2}.
\eea
This generating function admits the interpretation as the partition function of an infinite
tower of simple harmonic oscillators and the sequence of coefficients is recorded in the Online Encyclopaedia Of Integer sequences (OEIS) as the sequence A000714 \cite{OEISA000714}. It is close to, but not identical to, the partition function of a non-chiral scalar field in two dimensions, which has previously been discussed in the context of large $N$ gauge theory 
(see  for example equation (3.3) in \cite{Douglas1993}).  This reveals an unexpected bridge between the
combinatorics of non-semisimple walled Brauer algebras and a familiar physical structure
from oscillator systems associated with two-dimensional quantum field theory. 

Finally, we discuss potential future directions and applications of this work. The results are
relevant for the construction of orthogonal bases of matrix invariants in gauge theory and
AdS/CFT, where the walled Brauer algebra controls polynomial invariants of  complex matrices.  They also connect naturally with quantum
information theory, where mixed Schur--Weyl duality and walled Brauer algebras arise in
problems such as port-based teleportation and related symmetry-adapted protocols. We end
with a brief discussion of further directions, including the explicit construction of matrix
units in the non-semisimple regime and deeper structural understanding of the oscillator
interpretation. 

\section{Schur--Weyl duality and the walled Brauer algebra}
\label{sec:SWduality} 


In this section we briefly review the representation-theoretic framework underlying the
problem, namely Schur--Weyl duality for mixed tensor spaces and the role of the walled
Brauer algebra $B_N(m,n)$ as the commutant of the $U(N)$ action. We recall the
classification of irreducible representations in terms of Brauer representation triples,
describe the associated Bratteli diagrams in the semisimple regime $N\ge m+n$, and explain
how this picture is modified when $N<m+n$. This leads naturally to the dimension-correction
problem which motivates the restricted Bratteli diagrams introduced in \cite{SHOWBA}.

The representation theory of the unitary group $U(N)$ has widespread applications in
gauge-string duality and quantum information theory. $U(N)$ arises as the gauge group in
many important instances of gauge-string duality. In quantum information theory, and more
generally in quantum mechanics, it is the unitary group of transformations preserving the
inner product in $N$-dimensional Hilbert spaces.

A key result is Schur--Weyl duality, which gives the decomposition into irreducible
representations of the $n$-fold tensor power of the fundamental representation $V_N$. For
the case $ N \ge n $, this takes the simple form
\bea\label{largeNbasicSW}
V_N^{\otimes n}=\bigoplus_{Y\vdash n} V_Y^{U(N)}\otimes V_Y^{S_n}.
\eea
The direct sum is parametrised by Young diagrams with $n$ boxes, and each term in the sum
is a tensor product of an irrep $V_Y^{U(N)}$ of $U(N)$ with an irrep $V_Y^{S_n}$ of $S_n$.
This is an instance of the double commutant theorem, together with the fact that the algebra
of operators on $V_N^{\otimes n}$ which commutes with the $U(N)$ action given by
$U^{\otimes n}$ for group elements $U\in U(N)$, is the image of the group algebra
$\mC(S_n)$, where the symmetric group $S_n$ acts by permuting the factors of the tensor
product. Concretely, there is a change of basis from the tensor-space form to states
\bea
|i_1,\cdots,i_n\rangle \mapsto |Y,M_Y,m_Y\rangle,
\eea
with a Young diagram label $Y$, together with state labels $M_Y$ for the $U(N)$
representation associated with $Y$ and $m_Y$ for the $S_n$ representation associated with
$Y$.

The dimension $d_Y$ of the $S_n$ representation is given by the hook formula
\bea
d_Y &=& \frac{n!}{\prod_{a\in Y}\hbox{ hook lengths}(a)} \cr
&=& \frac{n!}{H(Y)},
\eea
where the product runs over all possible box positions $a$ within fixed Young diagram $Y$.
These dimensions give the multiplicities of the $U(N)$ representation in the decomposition
of tensor space. Specialising to $n=3$, the decomposition of tensor space in terms of
$U(N)$ irreps is
\bea\label{multiplUN}
V_N^{\otimes 3}=V_{\tiny{\yng(3)}}\oplus 2V_{\tiny{\yng(2,1)}}\oplus V_{\tiny{\yng(1,1,1)}}.
\eea
The symmetric and anti-symmetric irreps of $S_3$ have dimension $1$, while the mixed
representation has dimension $2$, as follows from the hook formula:
\bea
\begin{ytableau}
3 & 1 \\
1
\end{ytableau}
\qquad ; \qquad
\frac{3!}{3\times 1\times 1}=2.
\eea
Note the correlation between symmetric group dimensions and the multiplicities $(1,2,1)$ in
\eqref{multiplUN}, which is a direct consequence of the Schur--Weyl duality equation \eqref{largeNbasicSW}.

For the case $ n > N$, there is a simple cut-off on the height of the Young diagrams, leading
to the modification
\bea\label{finiteNBasicSW}
V_N^{\otimes n}=\bigoplus_{\substack{\hght(Y)\le N}} V_Y^{U(N)}\otimes V_Y^{S_n},
\eea
where the symbol $\hght(Y)$ denotes the height of the Young diagram $Y$, or equivalently the length of its first column.

In the mixed tensor case, for $N\ge (m+n)$, there is a generalisation of Schur--Weyl duality:
\bea\label{eq:stablemixedSW}
\Vmn=\bigoplus_{\gamma\in \BRT(m,n)} V^{U(N)}_{\gamma}\otimes V^{B_N(m,n)}_{\gamma},
\eea
where the set of Brauer representation triples $\BRT(m,n)$ consists of
\bea
&& \gamma=(k,\gamma_+,\gamma_-) \cr
&& 0\le k\le \min(m,n) \cr
&& \gamma_+ \hbox{ is a Young diagram with } (m-k) \hbox{ boxes} \cr
&& \gamma_- \hbox{ is a Young diagram with } (n-k) \hbox{ boxes}.
\eea
Thus irreducible representations are labelled by an integer $k$ together with two ordinary
Young diagrams, one associated with the fundamental sector and one with the
anti-fundamental sector. It is convenient to think of this data as extending 
pure tensor Schur--Weyl duality using the additional contraction structure present in the
mixed tensor space.  We  denote  by $c_1(\gamma_+), c_1(\gamma_-)$ lengths of the first columns of Young diagrams $\gamma_+,\gamma_-$ respectively. Notice that $c_1(\gamma_+)+c_1(\gamma_-)=\operatorname{height}(\gamma)$.

\begin{figure}[t]
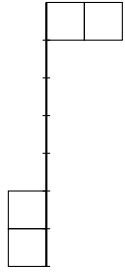

\centering
\MixedYoung{7}{0,0,0,0,0,1,1}{2,0,0,0,0,0,0}
\caption{Example of a mixed Young diagram for \(N=7\), associated with the Brauer
representation triple \((k=0,\gamma_+=(2),\gamma_-=(1,1))\).}
\label{fig:mixed-young-example}
\end{figure}

The triple $\gamma=(k,\gamma_+,\gamma_-)$ also determines a mixed Young diagram
$\Gamma(\gamma,N)$, which specifies the highest weight of the corresponding
$U(N)$ irrep $V^{U(N)}_{\gamma}$. The row lengths of the mixed Young diagram are
\bea\label{mixedYD}
&& R_i(\gamma)=r_i(\gamma_+) \qquad \hbox{ for } \qquad 1\le i\le c_1(\gamma_+) \cr
&& R_{N-i+1}(\gamma)=-r_i(\gamma_-) \qquad \hbox{ for } \qquad 1\le i\le c_1(\gamma_-).
\eea
In Figure~\ref{fig:mixed-young-example}, for $N=7$, we show the mixed Young diagram
associated with the Brauer representation triple
\(\gamma=(k,\gamma_+,\gamma_-)=(0,(2),(1,1))\).

In the range $N\ge (m+n)$, the dimension $\dim(V_{\gamma}^{B_N(m,n)})$ is independent of
$N$:
\bea\label{stabrangeDim}
\dim\bigl(V_{\gamma}^{B_N(m,n)}\bigr)=d_{m,n}(\gamma)=
\frac{m!\,n!}{k!\,H(\gamma_+)\,H(\gamma_-)}.
\eea
These dimensions give the multiplicities of the $U(N)$ irreps $V^{U(N)}_{\gamma}$ in tensor
space. A useful feature of the regime $N\ge m+n$ is that the representation-theoretic data
are stable as $N$ varies: in particular, the dimensions $d_{m,n}(\gamma)$ do not depend on
$N$. This large-$N$ stability provides the starting point for our later analysis of the
non-semisimple regime, where the corresponding multiplicities may deviate from the stable
formula.

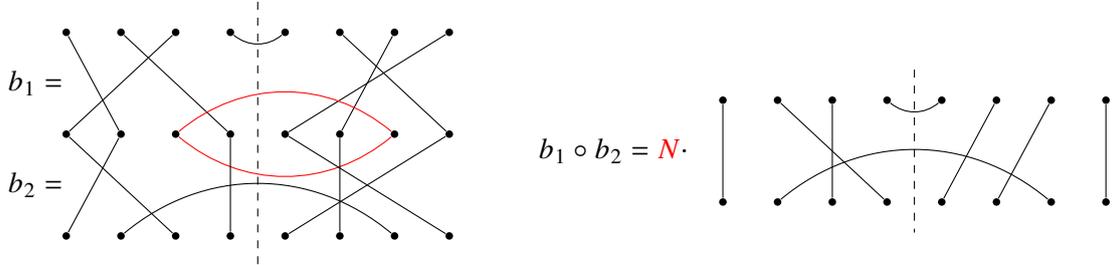
\begin{figure}[h!]
\centering
\begin{tikzpicture}[scale=0.9, every node/.style={inner sep=2pt}]
  \def\sep{0.8}
  \def\h{1.5}

  \foreach \i in {1,...,8} {
    \node[circle, fill=black, inner sep=1pt] (s\i) at ({\i*\sep}, {2*\h}) {};
    \node[circle, fill=black, inner sep=1pt] (sp\i) at ({\i*\sep}, {\h}) {};
  }

  \foreach \i in {1,...,8} {
    \node[circle, fill=black, inner sep=1pt] (pp\i) at ({\i*\sep}, {0}) {};
  }

  \node at (0.5*\sep, 1.5*\h) {$b_1=\ $};
  \node at (0.5*\sep, 0.5*\h) {$b_2=\ $};

  \draw[dashed] (4.5*\sep, {2.3*\h}) -- (4.5*\sep, {-0.3*\h});

  \draw (s1) -- (sp2);
  \draw (s2) -- (sp4);
  \draw (s3) -- (sp1);
  \draw (s4) to[bend right=40] (s5);
  \draw (s6) -- (sp8);
  \draw (s7) -- (sp6);
  \draw (s8) -- (sp5);
  \draw[red] (sp3) to[bend left=40] (sp7);

  \draw (sp1) -- (pp3);
  \draw (sp2) -- (pp1);
  \draw[red] (sp3) to[bend right=40] (sp7);
  \draw (sp4) -- (pp4);
  \draw (sp5) -- (pp8);
  \draw (sp6) -- (pp6);
  \draw (sp8) -- (pp5);
  \draw (pp2) to[bend left=40] (pp7);

  \def\compShift{-1.0}

\foreach \i in {1,...,8} {
  \node[circle, fill=black, inner sep=1pt] (c\i) at ({(\i+12)*\sep}, {2*\h+\compShift}) {};
  \node[circle, fill=black, inner sep=1pt] (cp\i) at ({(\i+12)*\sep}, {\h+\compShift}) {};
}

  \draw[dashed] (16.5*\sep, {2.3*\h+\compShift}) -- (16.5*\sep, {0.7*\h+\compShift});
  \node at (11.0*\sep, 1.5*\h+\compShift)
    {$b_1\circ b_2=\textcolor{red}{N}\cdot$};

  \draw (c1) -- (cp1);
  \draw (c2) -- (cp4);
  \draw (c3) -- (cp3);
  \draw (c4) to[bend right=40] (c5);
  \draw (c6) -- (cp5);
  \draw (c7) -- (cp6);
  \draw (c8) -- (cp8);
  \draw (cp2) to[bend left=40] (cp7);

\end{tikzpicture}
\caption{Example of graphical composition of two diagrams
$b_1,b_2\in B_N(4,4)$. The closed loop highlighted in red contributes a factor of
$N\in\mathbb{C}$, and the resulting composition again defines an element of
$B_N(4,4)$.}
\label{fig:WBA}
\end{figure}

The diagrams of the algebra $B_N(m,n)$ can be represented on mixed tensor
space by interpreting the lines in Brauer diagrams as contractions of tensor indices. This
gives a homomorphism
\[
\rho_{N,m,n}:B_N(m,n)\to {\rm End}\!\left(V_N^{\otimes m}\otimes \overline{V}_N^{\otimes n}\right).
\]
In the stable regime $N\ge m+n$, this map is faithful and its image is precisely the commutant of the action
of \(U^{\otimes m}\otimes \overline{U}^{\otimes n}\). Equivalently, the subalgebra of
\({\rm End}(\Vmn)\) commuting with the unitary action is the Brauer algebra \(B_N(m,n)\)
itself. In particular, \(B_N(m,n)\) is semisimple in this regime, i.e. it has a
non-degenerate trace form. Figure~\ref{fig:WBA} illustrates diagrammatic multiplication in the
Brauer algebra.

In the regime $N<(m+n)$, the situation changes substantially. The algebra $B_N(m,n)$ is no
longer semisimple, and the map \(\rho_{N,m,n}\) develops a non-trivial kernel. The quotient
of $B_N(m,n)$ by this kernel is a semisimple algebra \(\widehat{B}_{m,n}(N)\), which is
isomorphic to the commutant of \(U^{\otimes m}\otimes \overline{U}^{\otimes n}\) in the
mixed tensor representation. There is now a Schur--Weyl duality between $U(N)$ and
\(\widehat{B}_{m,n}(N)\):
\bea\label{eq:mixedSW}
V_N^{\otimes m}\otimes \overline{V}_N^{\otimes n}
=
\bigoplus_{\substack{\gamma\in \BRT(m,n)\\ \hght(\gamma)=c_1(\gamma_+)+c_1(\gamma_-)\le N}}
V_{\gamma}^{U(N)}\otimes V_{\gamma}^{\widehat{B}_N(m,n)}.
\eea
As in ordinary Schur--Weyl duality, there is a cut-off on \(\gamma\), now expressed by the
finite-\(N\) constraint \(\hght(\gamma)\le N\). But the non-semisimplicity of \(B_N(m,n)\)
has an additional consequence: the multiplicities of \(V_{\gamma}^{U(N)}\), or equivalently
the dimensions of the representations of the dual algebra \(\widehat{B}_N(m,n)\), may
differ from the stable large-\(N\) values. More precisely, they take the form
\bea\label{Corrdims}
d_{m,n,N}=\dim\bigl(V_{\gamma}^{\widehat{B}_N(m,n)}\bigr)=d_{m,n}(\gamma)-\delta_{m,n,N}(\gamma),
\eea
where \(\delta_{m,n,N}(\gamma)\ge 0\) measures the deviation from the stable formula
\eqref{stabrangeDim}. These corrections admit a description in terms of Bratteli diagrams
(\cite{StollWerth} and related references), but general closed formulae are not known.

This is the point at which the restricted Bratteli diagrams introduced in \cite{SHOWBA}
become useful. They provide a simplified combinatorial framework which isolates precisely
the data responsible for the correction terms \(\delta_{m,n,N}(\gamma)\), while taking as
input the stable-regime information encoded by \eqref{stabrangeDim}. In the regime
\(N=m+n-l\), the resulting diagrams display a new stability property: for
\(m,n\ge 2l-3\), their structure depends only on \(l\). This makes it possible to analyse
the correction problem in a universal way for fixed \(l\).

\section{Bratteli diagrams}

Bratteli diagrams provide a convenient combinatorial tool for organising the representation
theory of the walled Brauer algebra \(B_N(m,n)\) in the stable semisimple regime
\(N\ge m+n\). In this regime they encode the branching structure of Brauer representation
triples and allow one to compute irreducible dimensions by path counting. We briefly review
this construction and then explain how it is modified in the non-semisimple regime
\(N<m+n\). 

For a fixed pair \((m,n)\), the Bratteli diagram is a layered graph whose levels are labelled
by integers \(0\le L\le m+n\). The top level \(L=0\) consists of a single root node. For
\(1\le L\le m\), the nodes at level \(L\) are Brauer triples for the pair \((L,0)\), i.e. elements of the set  $ \BRT (L,0) $. For
\(m+1\le L\le m+n\), the nodes at level \(L\) are elements of 
\( \BRT (m,L-m)\). In particular, the last level \(L=m+n\) is naturally identified with the full
set \(\BRT(m,n)\). Edges connect nodes in adjacent levels whenever the corresponding triples
are related by a {\it Bratteli move}. Concretely, for \(L\le m\) one adds a box to
\(\gamma_+\) so as to obtain a valid Young diagram, while for \(L>m\) one either adds a box
to \(\gamma_-\) or removes a box from \(\gamma_+\). 

\begin{figure}[!ht]
\centering
\includegraphics[width=0.5\textwidth]{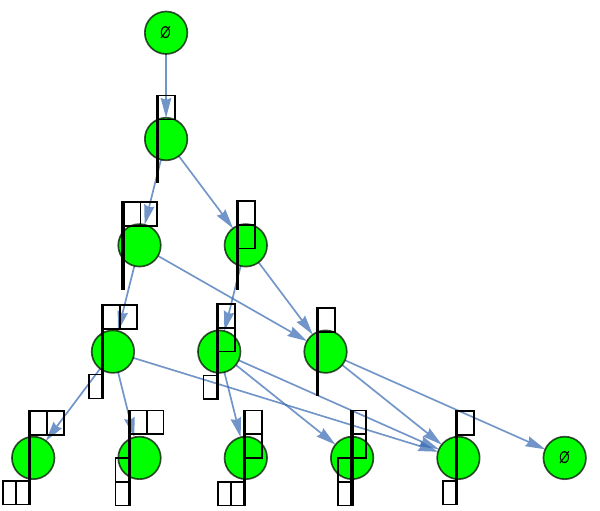}
\caption{Bratteli diagram for \(B_N(2,2)\) in the stable regime \(N\ge 4\).}
\label{B22large}
\end{figure}

Figure~\ref{B22large} shows the Bratteli diagram for \(B_N(2,2)\) in the stable regime
\(N\ge 4\). The sequence of layers corresponds to 
\bea
\BRT(0,0)\rightarrow \BRT(1,0)\rightarrow \BRT(2,0)\rightarrow \BRT(2,1)\rightarrow \BRT(2,2).
\eea
The nodes in the bottom layer correspond to the Brauer triples for \((m,n)=(2,2)\), while
the arrows encode the allowed Bratteli moves between consecutive levels.

A basic property of the Bratteli diagram is that the number of paths from the root at
level \(L=0\) to a given \(\gamma\in \BRT(m,n)\) is equal to the stable-regime dimension
\(d_{m,n}(\gamma)\) from \eqref{stabrangeDim}. Thus the dimension formula admits a direct
combinatorial interpretation in terms of path counting. For the example in
Figure~\ref{B22large}, one finds a single path for
\(\gamma=(k=0,\gamma_+=(2),\gamma_-=(1,1))\), while for
\(\gamma=(k=1,\gamma_+=(1),\gamma_-=(1))\) there are four paths, in agreement with
\eqref{stabrangeDim}. 

\begin{figure}[!ht]
\begin{center}
\includegraphics[width=0.5\textwidth]{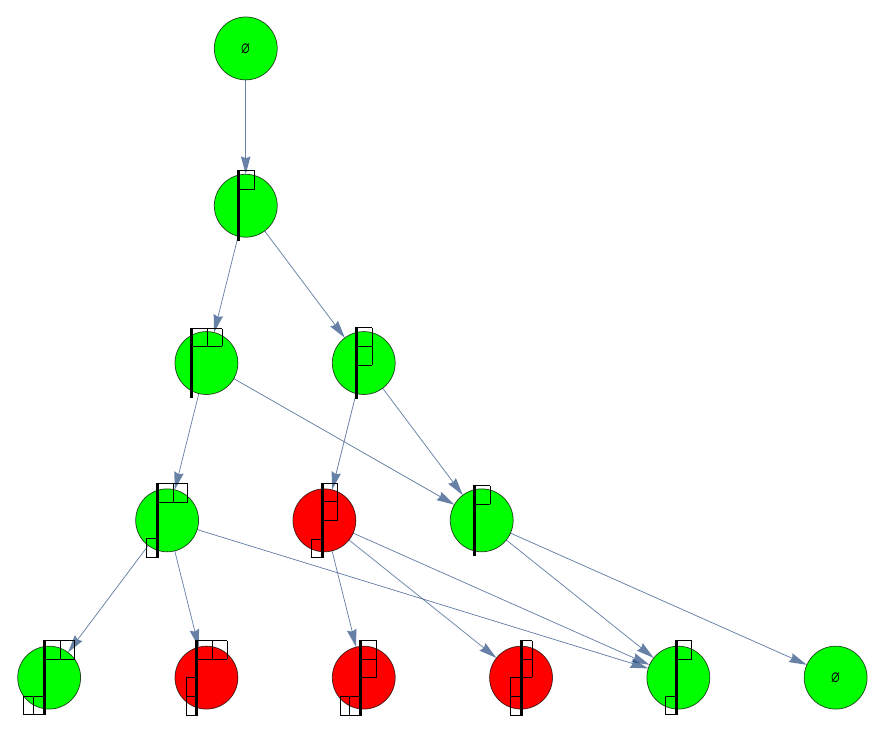}
\end{center}
\caption{Coloured Bratteli diagram for \(B_N(2,2)\) with \(N=2\). Red nodes violate the
finite-\(N\) constraint, while green nodes correspond to admissible Brauer triples.}
\label{B22Neq2}
\end{figure}

When \(N<m+n\), the same underlying layered graph may still be used, but now one must keep
track of the finite-\(N\) constraint
\[
\hght(\gamma)=c_1(\gamma_+)+c_1(\gamma_-)\le N.
\]
This leads to the notion of a {\it coloured Bratteli diagram (CBD)}: nodes violating the finite-\(N\)
constraint are coloured red, while the remaining nodes are coloured green. The red nodes
represent Brauer triples which are excluded in the Schur--Weyl decomposition at the given
value of \(N\), whereas the green nodes correspond to admissible triples. Figure~\ref{B22Neq2}
shows the coloured Bratteli diagram for \(B_{N=2}(2,2)\). 

Irreducible representations of the quotient algebra \(\widehat{B}_N(m,n)\) correspond to
the green nodes in the last layer. However, unlike in the stable regime, their dimensions
are no longer given simply by counting all paths from the root. The reason is that some of
these paths pass through red nodes at earlier levels, and such paths must be excluded. This
motivates the notion of an {\it admissible path}: a path from the root to a green node in the
final layer is called admissible if it does not pass through any red node at intermediate
levels. The corrected dimension \(\widehat d_{m,n,N}(\gamma)\) is then given by the number of
admissible paths ending at \(\gamma\). Equivalently,
\(\widehat d_{m,n,N}(\gamma)=d_{m,n}(\gamma)-\delta_{m,n,N}(\gamma)\), where
\(\delta_{m,n,N}(\gamma)\) counts the inadmissible paths, i.e. those passing through at least
one red node. 

In the example of Figure~\ref{B22Neq2}, the final-layer triple
\(\gamma=(k=1,\gamma_+=(1),\gamma_-=(1))\) has stable dimension \(d_{2,2}(\gamma)=4\), but
one of the four paths reaches it through a red node. Hence only three paths are admissible,
and therefore
\bea
\widehat d_{2,2,2}(\gamma)=4-1=3.
\eea
This simple example illustrates the general mechanism by which the non-semisimple regime
modifies the stable large-\(N\) counting. The main purpose of the next section is to explain
how the restricted Bratteli diagrams introduced in \cite{SHOWBA} isolate precisely the part
of the coloured diagram responsible for these corrections, leading to a more efficient
description of the modified dimensions.

\section{Restricted Bratteli diagrams and a new stability for $N$ close to $m+n$}

The coloured
Bratteli diagrams introduced in the previous section contain combinatorial
information which is also  encoded in the dimension formula \eqref{stabrangeDim}  from the stable regime. 
In \cite{SHOWBA}, this
motivated the introduction of \emph{restricted Bratteli diagrams} (RBD), which retain only
the part of the coloured diagram responsible for the dimension corrections. Their main
advantage is that they allow a computation of the dimension corrections in \eqref{Corrdims} while continuing
to use the stable-regime dimensions as input.

The restricted Bratteli diagram is obtained from the coloured Bratteli diagram by the
following reduction procedure:
\begin{itemize}
\item in the final layer, one keeps only those admissible Brauer triples (green nodes) whose dimensions
are modified;
\item in the earlier layers, one keeps only those excluded (red) nodes which admit paths to these
final-layer nodes;
\item one also retains the intermediate admissible nodes that lie on such paths.
\end{itemize}
In this way, the RBD is a smaller layered graph containing precisely the combinatorial data
needed to compute the correction terms in \eqref{Corrdims}. 

\begin{figure}[!ht]
\begin{center}
\includegraphics[width=0.08\textwidth]{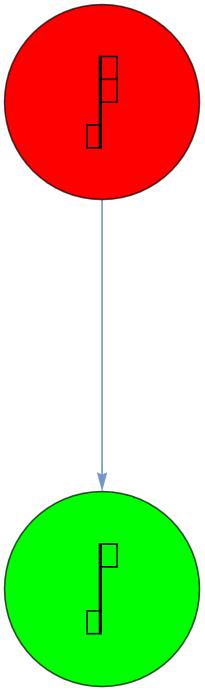}
\end{center}
\caption{Restricted Bratteli diagram for \(B_{N=2}(2,2)\).}
\label{RBD-B22Neq2}
\end{figure}

For the coloured Bratteli diagram of \(B_{N=2}(2,2)\), the corresponding restricted
Bratteli diagram is shown in Figure~\ref{RBD-B22Neq2}. In this example, the relevant
excluded node is
\[
\gamma=(k=0,\gamma_+=(1,1),\gamma_-=(1))\in \BRT(2,1),
\]
and its stable-regime dimension is
\bea
d_{2,1}(\gamma)=1.
\eea
This number is exactly the number of paths from the root to the excluded node in the
restricted diagram, and it gives the correction to the modified final-layer irrep of
\(B_{N=2}(2,2)\). Thus the RBD provides a particularly economical way of extracting the
dimension modification from the full coloured diagram. 

\begin{figure}[!ht]
	\begin{center}
		$B_{ N=1 } ( 2,2 )$  ~~~~  ~~~~ $ B_{ N=3 } ( 3,3 ) $  ~~~~  ~~~~  $ B_{ N=4 } ( 3,4 ) $  \\ 
		\includegraphics[scale=0.3]{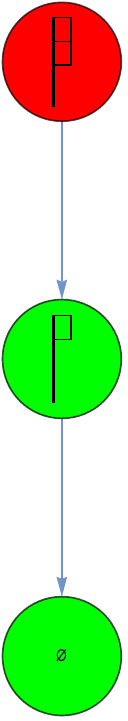} ~~~~~~~~~~~~ \includegraphics[scale=0.3]{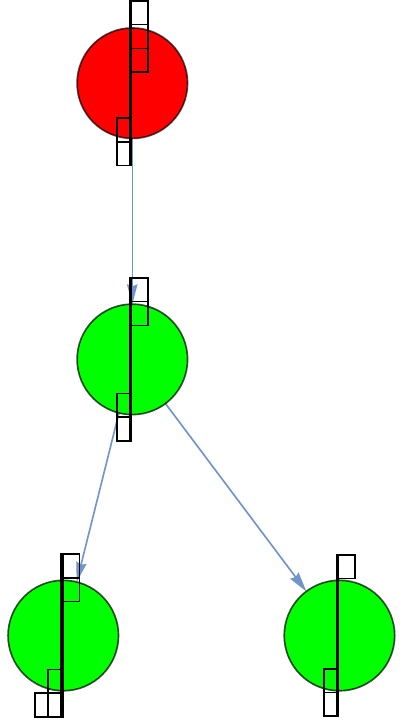} ~~~~~~~~~~  \includegraphics[scale=0.3]{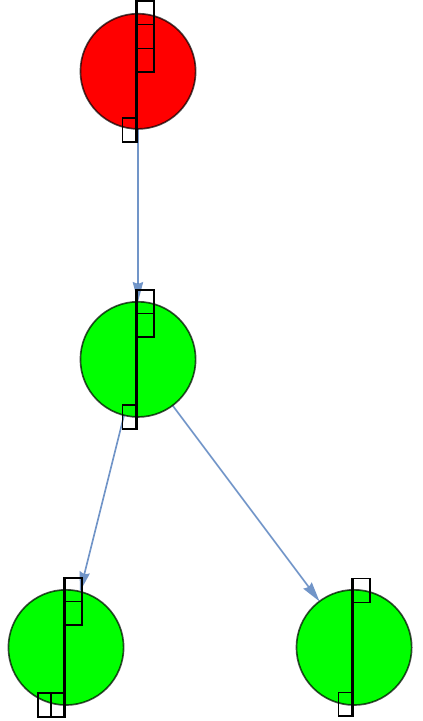}
	\end{center}
\caption{Stability for fixed $l = m + n - N  $ in the range  $m,n \ge ( 2l-3)$.  } 
\label{mnstabexamp}
\end{figure}

Using this characterisation, we implemented Mathematica code to generate the restricted
Bratteli diagrams for fixed \(m,n\) and \(N=m+n-l\), with \(l\ge 2\). The case \(l =1 \) is
trivial in the sense that no modified dimensions occur. Inspection of these examples led to
a remarkable stability phenomenon: for fixed \(l\), once \(m,n\ge 2l-3\), the restricted
Bratteli diagrams no longer depend on the individual values of \(m\) and \(n\).  This is illustrated in Figure~\ref{mnstabexamp} for the case $l=3$. The stable range here is $m,n \geq 3$. Once $m,n$ have reached these values, and are then increased in the progression from left to right, there are no further changes in RBD. We give a proof of this stability property for general $m,n,l$ in  \cite{SHOWBA}.

The structural properties of the RBD can be understood directly from the defining
constraints. Since we work with \(N=m+n-l\), an excluded node must satisfy
\[
c_1(\gamma_+)+c_1(\gamma_-) > m+n-l.
\]
It is therefore natural to define the \emph{excess height} \(\Delta\) by
\bea\label{defDeltxs}
c_1(\gamma_+)+c_1(\gamma_-)=(m+n-l)+\Delta.
\eea
By definition, one has \(\Delta\ge 1\) for every excluded node.

Now consider a node at depth \(d\), which corresponds to an irrep of \(B_N(m,n-d)\) with
label \((k,\gamma_+,\gamma_-)\). Writing \(|\gamma_\pm|\) for the numbers of boxes in the
Young diagrams \(\gamma_\pm\), we have
\bea\label{defk}
&& |\gamma_+|=m-k=c_1(\gamma_+)+|\gamma_+\setminus c_1|, \cr
&& |\gamma_-|=n-k-d=c_1(\gamma_-)+|\gamma_-\setminus c_1|.
\eea
Combining \eqref{defDeltxs} with \eqref{defk}, one obtains the basic relation
\begin{equation}\label{simpEq}
\boxed{
d+2k+|\gamma_+\setminus c_1|+|\gamma_-\setminus c_1|+\Delta=l.
}
\end{equation}
This identity has several immediate consequences. First, it shows that for fixed \(l\), all
the combinatorial data associated with excluded nodes are tightly constrained. Second, since
\(\Delta\ge 1\), it implies the bound
\bea
d\le l-1.
\eea
Hence the restricted Bratteli diagrams have maximal depth \(l-1\). This is one of the key
simplifications that makes the RBD effective: for fixed \(l\), only finitely many layers can
contribute, independent of how large \(m\) and \(n\) become. It is precisely this feature
that leads, in the next section, to a universal counting formulae and ultimately to the
oscillator generating function.

\section{Harmonic oscillator partition functions}

We now explain how the structural constraints derived in the previous section lead to a
universal oscillator partition function for restricted Bratteli diagrams. In the \((m,n)\)-stable regime,
the relevant combinatorics is governed by the depth \(d\), the excess height \(\Delta\), and
the partition data contained in the diagrams \(\gamma_+\setminus c_1\) and
\(\gamma_-\setminus c_1\). The key point is that, once \(l\) is fixed, the restriction
\eqref{simpEq}
severely limits the possible red nodes that can occur in the restricted Bratteli diagram.

A further consequence of the Bratteli moves is that the excess height cannot grow
arbitrarily at fixed depth. More precisely,
\bea\label{UPD}
\Delta(d)\le \min(l-d,d).
\eea
This inequality makes it possible to count red nodes at fixed \(l\) and \(d\) in terms of
partitions. Denoting by \(\mathcal R(l,d)\) the number of red nodes at depth \(d\), one
obtains the general formula
\bea\label{Redcount1}
\mathcal R(l,d)
=
\sum_{\Delta=1}^{\min(l-d,d)}
\;
\sum_{k=0}^{\left\lfloor \frac{l-d-\Delta}{2}\right\rfloor}
\;
\sum_{l_1=0}^{\,l-d-2k-\Delta}
p(l_1)\,p(l-d-2k-\Delta-l_1),
\eea
where \(p(r)\) denotes the number of partitions of the integer $r$. Thus the counting problem reduces to a finite
sum over the allowed values of \(\Delta\), \(k\), and the partition sizes carried by the
truncated diagrams. 

A particularly striking simplification occurs when one considers the form of $ \cR( l , d )$ for $ d$ small and  separately 
for $ d $ close to $l$.  The surprisingly simple result is that the counting in these two regimes can be related to  universal generating  function
\bea\label{Zuniv}
\mathcal{ Z}_{\rm univ}(x)=\sum_{s=0}^{\infty} x^s\,\mathcal Z(s)
=
\frac{x}{(1-x)(1-x^2)}\prod_{i=1}^{\infty}\frac{1}{(1-x^i)^2}.
\eea
As shown in the long paper~\cite{SHOWBA}, this is the generating function recorded in OEIS as A000714~\cite{OEISA000714},
and it admits a natural oscillator interpretation: the product factor
\(\prod_{i\ge 1}(1-x^i)^{-2}\) is the partition function of two infinite towers of harmonic
oscillators, while the additional prefactor \(x/[(1-x)(1-x^2)]\) corresponds to extra modes
of energies \(1\) and \(2\). In this sense, \(\mathcal Z_{\rm univ}(x)\) is the partition
function of an infinite collection of simple harmonic oscillators. 

 For \(d\) close to \(l\), and also for sufficiently small \(d\), the bound
\(\Delta\le \min(l-d,d)\) simplifies, leading to the explicit expressions
\begin{equation}\label{RedsOscSum}
\boxed{
\mathcal{R}(l,d)=
\left\{
\begin{array}{ll}
\mathcal{Z}_{\text{univ}}(l-d)-\mathcal{Z}_{\text{univ}}(l-2d),
& \text{for } 1\le d\le \left\lfloor \frac{l}{2}\right\rfloor,\\[0.2cm]
\mathcal{Z}_{\text{univ}}(l-d),
& \text{for } \left\lceil \frac{l}{2}\right\rceil\le d\le l-1.
\end{array}
\right.
}
\end{equation}
Thus, in the \((m,n)\)-stable regime, the distribution of red nodes in the restricted
Bratteli diagram is governed by a single universal sequence. This is one of the main
results  of the analysis.

The same generating function also governs the counting of green nodes.  Let
\(\mathcal G(l,d)\) denote the number of green nodes in the restricted Bratteli diagram at
depth \(d\). The key observation is that there is a bijection between the green nodes at
depth \(d=0\) and the red nodes with \(\Delta=1\) at higher depths. More precisely, every
green node has a unique minimal-depth red ancestor, and this ancestor must satisfy
\(\Delta=1\). This yields a bijection between the two sets and hence identifies the green
counting problem with a distinguished part of the red counting problem. The green nodes of the RBD at $ d=0$ 
 are the Brauer representation triples which receive non-zero dimension corrections. 

Using the bijection with the  $ \Delta=1$ red nodes, we find
\bea\label{GreensUniv}
\boxed{
\mathcal G(l,0)=\mathcal Z_{\rm univ}(l-1).
}
\eea
In fact, the same reasoning extends to higher depths and gives
\[
\mathcal G(l,d)=\mathcal Z_{\rm univ}(l-d-1).
\]
For the purposes of the present proceedings note, the main message is that both the red-node
and green-node counting problems are controlled by the same universal oscillator partition
function.

This oscillator structure is not only combinatorially elegant, but also physically suggestive.
In the matrix-model and AdS/CFT setting, the quotient algebra \(\widehat{B}_N(m,n)\)
controls orthogonal bases of polynomial matrix invariants for a complex matrix $Z$ of size $N$, which transforms under the adjoint representation of $U(N)$ \cite{KimuraRamgoolam2007} (this application also extends to multi-complex matrices \cite{Kimura:2009jf}). These polynomial invariants are generated by traces of products involving  $Z$ and the hermitian conjugate $ Z^{\dagger}$.  The integers $m,n$ count the degree in $Z$ and $Z^{ \dagger} $ respectively. The orthogonal bases are constructed from matrix units in the sub-algebra of $ \widehat B_N ( m , n )$ which commute with $S_m \times S_n$. These matrix units give the  Artin--Wedderburn decomposition of these permutation centralizer  algebras \cite{MattioliRamgoolam2016} and take the form: 
\bea
Q^{\gamma}_{r_1,r_2,\mu\nu}, 
\eea
where $ \gamma $ is a Brauer representation triple, $ r_1 , r_2$ are Young diagrams with $m$ and $n$ boxes respectively, 
and $\mu , \nu $ run over the multiplicities of reduction of the walled Brauer representation  $\gamma $ to
the representation  $ ( r_1,r_2)$ of the group algebra of $ S_m \times S_n$. 
Thus the  combinatorics that governs restricted Bratteli diagrams and dimension
corrections also feeds into the construction of orthogonal operator bases in gauge theory.
From this perspective, the appearance of the oscillator partition function suggests an
unexpected bridge between non-semisimple Brauer combinatorics and structures familiar from
large-\(N\) physics.

\normalsize

\section{ Summary and outlook}

The results of the paper \cite{SHOWBA}, and summarised in this proceedings contribution,  are motivated by applications of walled Brauer algebras in both
gauge--string duality and quantum information theory. In the AdS/CFT setting, these
algebras provide a natural framework for constructing orthogonal bases of general (not necessarily holomorphic) polynomial 
matrix observables from complex matrices $Z_a$, with basis labels from the representation theory of mixed tensor spaces.  Walled Brauer algebras and closely related centralizer algebras have been used in the study of exact multi-matrix correlators, orthogonal operator bases, the one-loop
dilatation operator in the quarter-BPS sector (Bogomolny–Prasad–Sommerfield), and the map between quarter-BPS operators
and bubbling geometries \cite{LLM} in the AdS dual \cite{KimuraRamgoolam2007,BhattacharyyaCollinsKoch2008,BhattacharyyaKochStephanou2008,BrownHeslopRamgoolam2008,BrownHeslopRamgoolam2009,KimuraRamgoolam2008,Kimura:2009jf,Kimura2010,KimuraLin2012,MattioliRamgoolam2016}. 

From the present perspective, the non-semisimple regime is especially interesting because
the quotient algebra \(\widehat{B}_N(m,n)\) controls the multiplicities appearing in mixed
Schur--Weyl duality, while the corresponding modified dimensions describe the breakdown of
the stable large-\(N\) formulas. In matrix-model language, these correction mechanisms are
therefore relevant for understanding finite-\(N\) effects in the organisation of
gauge-invariant operators and their orthogonal bases. The restricted Bratteli diagrams
developed here provide a tractable combinatorial framework for these finite-\(N\)
corrections, and the emergence of the universal oscillator generating function suggests that
there may be a deeper algebraic structure underlying the quotient algebra itself.

On the quantum-information side, walled Brauer algebras arise naturally in mixed
Schur--Weyl duality and in the algebra of partially transposed permutation operators. Their
efficient implementation in quantum circuits has led to applications in quantum
transmission protocols and, in particular, port-based teleportation
\cite{StudzinskiMlynikMozrzymasHorodecki2025,GrinkoBurchardtOzols2023,Nguyen2023,
GrinkoBurchardtOzols2024,FeiTimmermanHayden2023,WillsHsiehStrelchuk2024,Grinko2025,
StudzinskiStrelchukMozrzymasHorodecki2017,StudzinskiMozrzymasKopszak2021,
MozrzymasStudzinskiHorodecki2018}. In that setting, \(N\) is the dimension of a Hilbert
space \(\mathcal{H}\), Alice and Bob share \(m\) entangled pairs in \((\mathcal{H}\otimes \mathcal{H})^{\otimes m}\),
and these serve as a resource for teleporting states in \(\mathcal{H}^{\otimes n}\). The relevant
fidelity expressions can be written in terms of traces of elements of \(B_N(m,n)\), which
makes the representation theory of the walled Brauer algebra directly relevant for explicit
calculations. 

More broadly, the same algebraic structures have also appeared in higher-order quantum
operations, mixed Schur sampling, and symmetry reduction for optimisation problems with
unitary-equivariant constraints
\cite{QuintinoEbler2022,CerveroMartinMancinskaTheil2024,GrinkoOzols2024}. These examples
suggest that the non-semisimple representation theory of walled Brauer algebras may become
increasingly relevant as one seeks efficient symmetry-adapted descriptions of quantum
protocols and algorithms. 

The combinatorics of restricted Bratteli diagrams and their surprising relation to oscillator partition functions found in this work suggests a rich combinatorial mathematical physics of mixed tensor representations in the non-semisimple regime, with the full structure yet to be uncovered, and with potentially wide-ranging implications for gauge-string duality and quantum information theory. Specifically our main results show
that the problem of calculating corrected dimensions in the non-semisimple regime admits a surprisingly rigid organisation: the restricted Bratteli diagrams exhibit an \((m,n)\)-stability property, and both the red-node and
green-node counting problems are governed by the same universal generating function
\(\mathcal Z_{\rm univ}(x)\), with its interpretation in terms of an infinite tower of
simple harmonic oscillators. This strongly suggests that the non-semisimple combinatorics
of \(B_N(m,n)\) is controlled by structures that are more universal than might have been
expected from the original path-counting problem. 

A natural next direction is to move from counting results to explicit representation-theoretic
constructions. In particular, an important open problem is to develop practical algorithms
for matrix units and Artin--Wedderburn-type bases in the non-semisimple regime, making the
connection between the abstract quotient algebra and its matrix realisation more explicit.
Another direction is to extend the present analysis beyond the small-\(l\) regime and to
understand whether the oscillator structure persists in a broader form. It is also tempting
to ask whether similar universal generating functions appear for related diagram algebras,
mixed-tensor dualities, or other symmetry-restricted algebras arising in physics.

The  interplay between stable large-\(N\) representation theory,
finite-\(N\) corrections, restricted Bratteli diagrams, and oscillator-type generating
functions promises to be of interest not only for the applications discussed above, but also as a
structural bridge between representation theory and mathematical physics with the full ramifications to be found. 

\normalsize

\begin{center}
{\bf \Large Acknowledgments}
\end{center} 

We are grateful to the organizers of the Corfu workshop on ``Cartan, Generalised and Noncommutative Geometries, Lie Theory and Integrable Systems Meet Vision and Physical Models'' in the context of the ``Cost Action CaLISTA General Meeting 2025''
and the Corfu Summer Institute 2025, for the invitation to present this work and contribute to the proceedings. 
 SR is supported by the Science and Technology Facilities Council (STFC) Consolidated Grant ST/T000686/1 
``Amplitudes, strings and duality''. SR is grateful for a Visiting Professorship at the Dublin Institute for Advanced Studies,  held during 2024, when the project leading to \cite{SHOWBA} was initiated. 
M.S. acknowledges support from the Polish National Science Centre (NCN) under SONATA BIS grant no. UMO-2024/54/E/ST2/00316, which supported the preparation of this manuscript.


\end{document}